\documentclass[aps,prl,reprint,showpacs,nofootinbib,superscriptaddress]{revtex4-1}
\usepackage{xcolor}
\usepackage{comment}
\usepackage{hyperref}
\usepackage{graphicx}
\usepackage{epsfig}
\usepackage{bm}
\usepackage{color}
\usepackage{float}
\usepackage{dcolumn}
\usepackage{multirow} 
\usepackage{hyperref}
\usepackage{amsmath}
\usepackage{amsfonts}
\usepackage{mathtools}

\definecolor{red}{rgb}{1,0.,0}
\begin{document}
\title{Density matrix based perturbative corrections for improved quantum simulation accuracy}

\author{T. D. Morris}
\affiliation{Physics Division, \ Oak\ Ridge\ National\ Laboratory,\
  Oak\ Ridge,\ TN,\ 37831,\ USA}
\affiliation{Quantum Computing Institute,\ Oak\ Ridge\ National\ Laboratory,\
  Oak\ Ridge,\ TN,\ 37831,\ USA}
  
\author{Z. P. Parks}
\affiliation{Computer Science and Mathematics Division,\ Oak\ Ridge\ National\ Laboratory,\ Oak\ Ridge,\ TN,\ 37831,\ USA}
\affiliation{Quantum Computing Institute,\ Oak\ Ridge\ National\ Laboratory,\
  Oak\ Ridge,\ TN,\ 37831,\ USA}
  
\author{A. J. McCaskey}
\affiliation{Computer Science and Mathematics Division,\ Oak\ Ridge\ National\ Laboratory,\ Oak\ Ridge,\ TN,\ 37831,\ USA}
\affiliation{Quantum Computing Institute,\ Oak\ Ridge\ National\ Laboratory,\
  Oak\ Ridge,\ TN,\ 37831,\ USA}
  
\author{J. Jakowski}
\affiliation{Computational Sciences and Engineering Division,\ Oak\ Ridge\ National\ Laboratory,\
Oak\ Ridge,\ TN,\ 37831,\ USA}
\affiliation{Quantum Computing Institute,\ Oak\ Ridge\ National\ Laboratory,\
  Oak\ Ridge,\ TN,\ 37831,\ USA}
  
\author{R. C. Pooser}
\affiliation{Computational Sciences and Engineering Division,\ Oak\ Ridge\ National\ Laboratory,\ Oak\ Ridge,\ TN,\ 37831,\ USA}
\affiliation{Quantum Computing Institute,\ Oak\ Ridge\ National\ Laboratory,\
  Oak\ Ridge,\ TN,\ 37831,\ USA}

\begin{abstract}
We present error mitigation (EM) techniques for noisy intermediate-scale quantum computers (QC) based on density matrix purification and perturbative corrections to the target energy. We incorporate this scheme into the variational quantum eigensolver (VQE) and demonstrate chemically-accurate ground state energy calculations of various alkali metal hydrides using IBM quantum computers. Both the density matrix purification improvements and the perturbative corrections require only meager classical computational resources, and are conducted exclusively as post-processing of the measured density matrix.  The improved density matrix leads to better simulation accuracy at each step of the variational optimization, resulting in a better input into the next optimization step without additional measurements. Adding perturbative corrections to the resulting energies further increases the accuracy, and decreases variation between consecutive measurements. These EM schemes allow for previously unavailable levels of accuracy over remote QC resources.
\end{abstract}

\maketitle

{\it Introduction.---}
Noisy intermediate scale quantum (NISQ) computers can be used to test small algorithms in support of codesign efforts to improve the performance of scientific applications. Error mitigation (EM) is a necessary step  for improving quantum computing results obtained from these devices. Recent studies have shown that NISQ devices can yield accurate results for a range of scientific simulation problems when EM is used to characterize systematic device noise and adjust expectation value data accordingly~\cite{kandala2017,kandala_error_2019,endo_practical_2018,temme_error_2017,li_efficient_2017}. 
In particular, quantum chemistry calculations have reached chemical accuracy for an array of small molecules with minimal basis sets~\cite{cao_quantum_2018,kandala_hardware-efficient_2017,Romero2018,romero_strategies_2017,yung_transistor_2014,peruzzo_variational_2014,hempel_quantum_2018,Ryabinkin2018,chembench}. 
EM has also been used to adjust resultant data in nuclear bound state calculations~\cite{Dumitrescu2018}, scalar field theory computations~\cite{yeter-aydeniz_scalar_2018}, and in Hamiltonian and quantum state learning applications~\cite{KAH_inprep}. 

Although it has been used in many applications involving the variational quantum eigensolver, EM is applicable to many other algorithms, including in hybrid machine learning applications that train a quantum register to perform state preparation, quantum approximate optimization~\cite{farhi_quantum_2014}, or in quantum imaginary time evolution~\cite{motta_quantum_nodate}. 
EM requires an assumption about the underlying error model, and the results depend on characterizing how well the machine approximates the model. We previously demonstrated that building assumptions about state preparation as well as the expected final density matrix into EM allows for higher accuracy~\cite{chembench, yeter-aydeniz_scalar_2018}. In this case EM is a characterization of how far the machine deviates from our assumptions. 

Here, we augment EM in two ways for quantum chemistry calculations. These modifications extend the reach of an array of molecules to chemical accuracy, even when programming a quantum computer remotely from the QASM level, where one has no gate or pulse level control (which would otherwise allow one to carry out more accurate gate and control-level characterization for use in EM~\cite{Eugene-mitigability}). 
First, we take into account symmetries of the problem and modify our density purification scheme, enforcing spin symmetry, to achieve a more physical density matrix. Second, we apply perturbative corrections which are motivated by interpreting unitary operation as acting on the Hamiltonian, as opposed to the wavefunction.  This interpretation is similar in spirit to the In-Medium Similarity Renormalization Group (IMSRG) theory~\cite{IMSRG-16}, which is a promising framework for solving the many-body problem.  
Within the IMSRG, the coupling of a subset of the Hilbert space to its complement is suppressed via a set of differential equations.  
When dealing with closed-shell ground states, this decreasing coupling enables computable corrections to energy. 
In this work, we demonstrate that an analagous perturbative correction, combined with energies arising from a symmetry-corrected, purified density matrix, can yield chemically accurate results for a wide range of molecules and system sizes.

{\it Molecular hamiltonians and basis sets.---}We start by considering a recently proposed ``benchmark'' class of molecules from the alkali-hydrides: H$_2$, LiH, NaH, KH  \cite{chembench}. Computation of ground state energy, correlation corrections, and potential energy curves  have become basic benchmark tasks in the NISQ computing era. The electronic structure calculations required to compute these phenomena correspond to a  Hamiltonian with a nuclear repulsion term, a one-electron term which accounts for electronic kinetic energy and electrodynamic interactions with the inner core, and a two electron term that accounts for electronic interactions:

\begin{eqnarray}
\label{eq:H_tot}
H&= & H_0 + \sum_{p,q} h_{pq} a_p^\dagger a_q   
         + \frac{1}{4} \sum_{p,q,r,s} g_{pqrs} a_p^\dagger a_q^\dagger a_s a_r,
\end{eqnarray}
where $H_0$ is the nuclear repulsion, $p$, $q$, $r$, $s$ index molecular spin orbitals,  $a_p^\dagger$,  $a_q$ etc.\ are electron creation and annihilation operators,  $h_{pq}$  are matrix elements of the core Hamiltonian, and $ g_{pqrs}$ are anti-symmetrized two-electron repulsion integrals $g_{pqrs} = \langle pq||rs\rangle$.
We freeze all but the highest occupied and lowest occupied orbitals that arise from a hartree-fock calculation~\cite{chembench}.  Additionally, for a given single Slater determinant $|\phi \rangle$,
one can always exactly rewrite the above hamiltonian in normal ordered form using Wick's theorem:
\begin{eqnarray}
\label{eq:H_N_tot}
H_N&= & E_0 + \sum_{p,q} f_{pq}:a_p^\dagger a_q:   
         + \frac{1}{4} \sum_{p,q,r,s} \Gamma_{pqrs} :a_p^\dagger a_q^\dagger a_s a_r:,\notag \\
\end{eqnarray}
where 
\begin{align}
E_0 &= H_0+\sum_i h_{ii}+\frac{1}{2}\sum_{ij}g_{ijij}\,, \\
f_{pq} &= h_{pq}+\sum_i g_{piqi} , \\
\Gamma_{pqrs} &= g_{pqrs} .
\end{align}  
Here $:a_p^\dagger \ldots a_q:$ indicates that the operator is normal-ordered, i.e. that $\langle \phi |:a_p^ \dagger \ldots a_q:|\phi\rangle=0$. Also here and for the remainder of this work,  i,j,k,$\ldots$ (a,b,c,$\ldots$) are indices of (un)occupied orbitals of the respective $|\phi\rangle$.

{\it Mapping to qubits.---} In order to map the problem onto qubits, we use the standard Jordan-Wigner tranformation~\cite{jordan_uber_1928}.
\begin{figure} 
\centering
\includegraphics[width=\columnwidth]{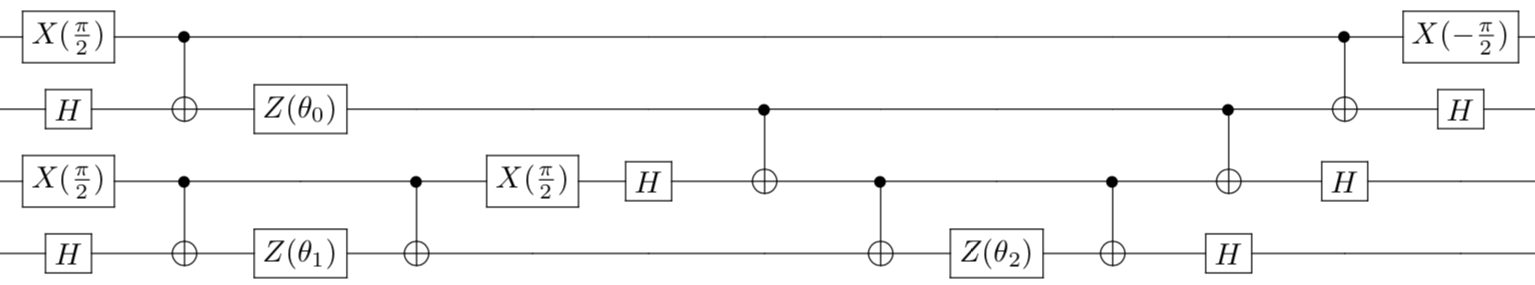}
\caption{The three parameter UCC insired ansatz for 4 qubits and 2 electrons. This ansatz consists of 3 parameters.  The parameter $\theta_0$ controls the double excitation from the doubly occupied lower spatial orbital into the doubly occupied higher orbital.  The parameter $\theta_1$  ($\theta_2$) controls single excitation amplitude within  $\alpha$ spin ($\beta$ spin) block.}
\label{fig:uccsdex}
\end{figure}
For our experiments, we employ the unitary coupled cluster-inspired ansatz found in \cite{chembench} and shown in Figure~\ref{fig:uccsdex}.  
Simulations show that this ansatz is capable of reproducing the exact energy of any  2 electron 4 spin-orbital singlet states, and is thus particularly suitable for the modeling of ground states of each alkali-hydride within the frozen core approximation.  We note that instead of tranforming the Hamiltonian to the Pauli representation, we instead use Pauli expectation values in order to construct the two body reduced density matrix, or 2-RDM, which is sufficient for evaluation of energy and key to our approach for error mitigation.

%
{\it Error Mitigation Techniques.---}Several methods for EM in near-term quantum hardware have been proposed including Richardson extrapolation \cite{Benjamin-PRX_17-EM,Gambetta-PRL_17-EM,Benjamin-PRX_18-EM}, 
the quasiprobability method \cite{Gambetta-PRL_17-EM,Benjamin-PRX_18-EM}, 
the quantum subspace expansion \cite{deJong-PRA_17}, 
and ancilla qubit stabilizers \cite{Benjamin-PRL-19-EM}. 
Employing quantum algorithms that exploit symmetries of spin  and conserve particle number during the VQE optimization leads to reduced search space, improved convergence and  decreased error rates since only a fraction of the Hilbert space is sampled. 
However, in practice, the spin and electron number are not exactly conserved due to unwanted cross-talk and systematic noise (over and under rotations).


Density matrix purification schemes are essentially methods of projecting inexact, unphysical, or ensemble density matrices onto the the density matrix of closest pure state, and can thus correct for noise that violates number or spin symmetry conservation.  We accomplish this via a procedure referred to in literature as McWeeney purification \cite{purif}, where iterative application of the matrix polynomial ${\cal P} \to 3{\cal P}^2-2{\cal P}^3$, drives eigenvalues that were originally close to 0 or 1 to exactly 0 or 1.  

Although this process is appropriate if one has the full A-body density matrix, as in our simulations, one can not appeal to this procedure for purifying a 2-RDM arising from a 3- or higher body system.  In this case, purification must proceed via semi-definite constraints subject to N-representability as pursued in \cite{marginal}. This procedure was previously exploited to dramatically improve final energies in benchmark simulations \cite{chembench}. To further improve the quality of the measured 2-RDM, we impose spin symmetry on the measured density matrix elements before subjecting the 2-RDM to purification.  This amounts to measuring only density matrix elements that conserve total spin; and in the 2 electron case only $S_z=0$ density matrix elements are allowed.  Further, we force spin reflection symmetry of the 2-RDM to be obeyed by averaging elements with their appropriate spin reflection, resulting in a measured density matrix that is more physical prior to purification.  We could also have only measured one spin reflection in order to cut down on measurements.  Here and for the proceeding perturbative correction, we propagate statistical error by performing simple bootstrapping, where each measured circuit is resampled 10,000 times; our error analysis proceeds after each resampling \cite{Efron1979}.  This procedure produces distribution average and errors that have been shown in previous work to be consistent with other forms of statistical error propagation \cite{chembench}.

{\it Perturbative correction.---} Attempting to apply perturbation theory on beyond-single slater determinant wavefunctions is a very complicated, if feasible process. To do so with the ansa\"etze typically employed in quantum computations would require so much classical computation as to likely make the benefit of quantum computing moot.
It is however possible to formulate an accurate approximation of perturbation theory, where one works not with the wavefunction as it evolves, but instead the hamiltonian itself.  We will take the unitary coupled cluster (UCC) wavefunction as a model, where the ground state is written as
\begin{equation}
    |\Psi\rangle = e^{T-T^\dagger}|\phi\rangle = U|\phi\rangle \,,
\end{equation}
with $|\phi\rangle$ typically the Hartree-Fock (HF) determinant, and the amplitudes of $T$ are varied to minimize the energy, which now takes the form
\begin{equation}\label{eq:energy_ucc}
    E = \langle\Psi|H|\Psi\rangle = \langle\phi|U^\dagger H U|\phi\rangle = \langle\phi|\bar{H}|\phi\rangle \,.
\end{equation}
It can be seen from Eq.~\ref{eq:energy_ucc}, it is valid to think of the transformation as acting on the Hamiltonian to produce $\bar{H}$, whose matrix elements connecting $|\phi\rangle$ to higher excitations are suppressed, leaving $|\phi\rangle$ as the exact eigenstate of $\bar{H}$.  This is the basic interpretation of how the IMSRG approaches the A-body diagonlization problem.  Working now with $\bar{H}$ allows one to appeal to the simple formula for perturbation theory,
\begin{equation}\label{eq:mp2_form}
    \Delta \bar{E}^{[2]} = \sum_{ia} \frac{|\bar{f}_{ia}|^2}{\bar{\epsilon}_i-\bar{\epsilon}_a}+ \frac{1}{4}\sum_{ijab}\frac{|\bar{\Gamma}_{ijab}|^2}{\bar{\epsilon}_i+\bar{\epsilon}_j-\bar{\epsilon}_a-\bar{\epsilon}_b} \,,
\end{equation}
where all matrix elements are those of the transformed Hamiltonian, normal ordered with respect to $|\phi\rangle$ (we define the matrix elements and the energy denominators in the supplemental information).  We have omitted, three- and higher- body pieces that could be induced
by the transformation, their importance and analogy to traditional coupled cluster perturbative triples and higher rank approximations will be explored in a future work.  As opposed to attempting to measure the matrix elements found in Eq.~\ref{eq:mp2_form} on hardware,which would be exceedingly expensive, we instead approximate them with our measured 2-RDM. We do so by noticing that the derivative with respect to a given cluster amplitude can be 
approximately written in the following ways,
\begin{equation}\label{eq:derive_general}
    \frac{\delta E}{\delta T_I} \approx 2\Re(\langle \phi | [A_I,\bar{H}]|\phi \rangle) \approx 2\Re(\langle\phi|e^{T^\dagger-T}[A_I,H]e^{T-T^\dagger}|\phi\rangle) \,,
\end{equation}
where $A_I$ are typical cluster type excitations out of the occupied Slater determinant $|\phi\rangle$. 
As the solution is asymptotically approached, both sides must be analytically zero and approximately equal.    
We observe that the RHS of Eq.~\ref{eq:derive_general} are nothing but the matrix elements encountered in the numerators of Eq.~\ref{eq:mp2_form}, whose expressions in terms of the bare hamiltonian and density matrix elements are presented in 
Eqs. S4,5, supplemental material~\cite{Supplemental}.
We also approximate the transformed energies $\bar{\epsilon}_i,\bar{\epsilon}_a$ using the measured 2-RDM by appealing to a connection between cluster amplitudes and certain density matrix elements (see Eqs. S6 and S7,~\cite{Supplemental}). 
The purified 2-RDM can then be used to form an estimate of $\Delta \bar{E}^{[2]}$, that when added to the purified energies, provides a more robust ground-state energy estimate, termed in this work as RDM-PT2.  Our correction depends only on the measured 2-RDM, and thus does not actually depend on using the UCC ansatz, which is corroborated by the fact that our simulated ansatz is not that of UCC theory.

All simulations on hardware consisted of 2-electrons correlated in the highest occupied and lowest unnoccupied orbitals, however, the RDM-PT2 correction is appropriate even for frozen orbitals.  It simplifies to simple M{\o}ller-Plosset perturbation theory when the 2-RDM is that of a single slater determinant, and becomes a better approximation as the ansatz becomes perturbatively close to the true ground state.
If the frozen orbitals represent only dynamical correlations, it is likely their contribution to the ground state energy can be captured perturbatively.  One caveat is that this necessitates approximations to the 3-RDM,
since $\bar{\Gamma}_{ijab}$ depends on the 3-RDM for more than 2-fermion systems (see Eq.~S7, \cite{Supplemental}).  Fortunately the 3-RDM elements can be approximated with a simple formula in terms of the 1- and 2- RDM matrix elements (Eq.~S8, \cite{Supplemental}), which is derived in \cite{mazziotti2007reduced}. This allows for accounting for all orbitals, even for our largest system KH, which consisted of 20 electrons in 28 orbitals.  This correction scales as ~o$^2$a$^2$v$^2$, where o is the number of frozen core orbitals, a is the number of active orbitals, and v is the number of frozen virtual orbitals. Thus our correction scales roughly the same as traditional CC theory truncated at singles and doubles.

\begin{figure}[hb]
    \centering
    \includegraphics[width=0.95\columnwidth]{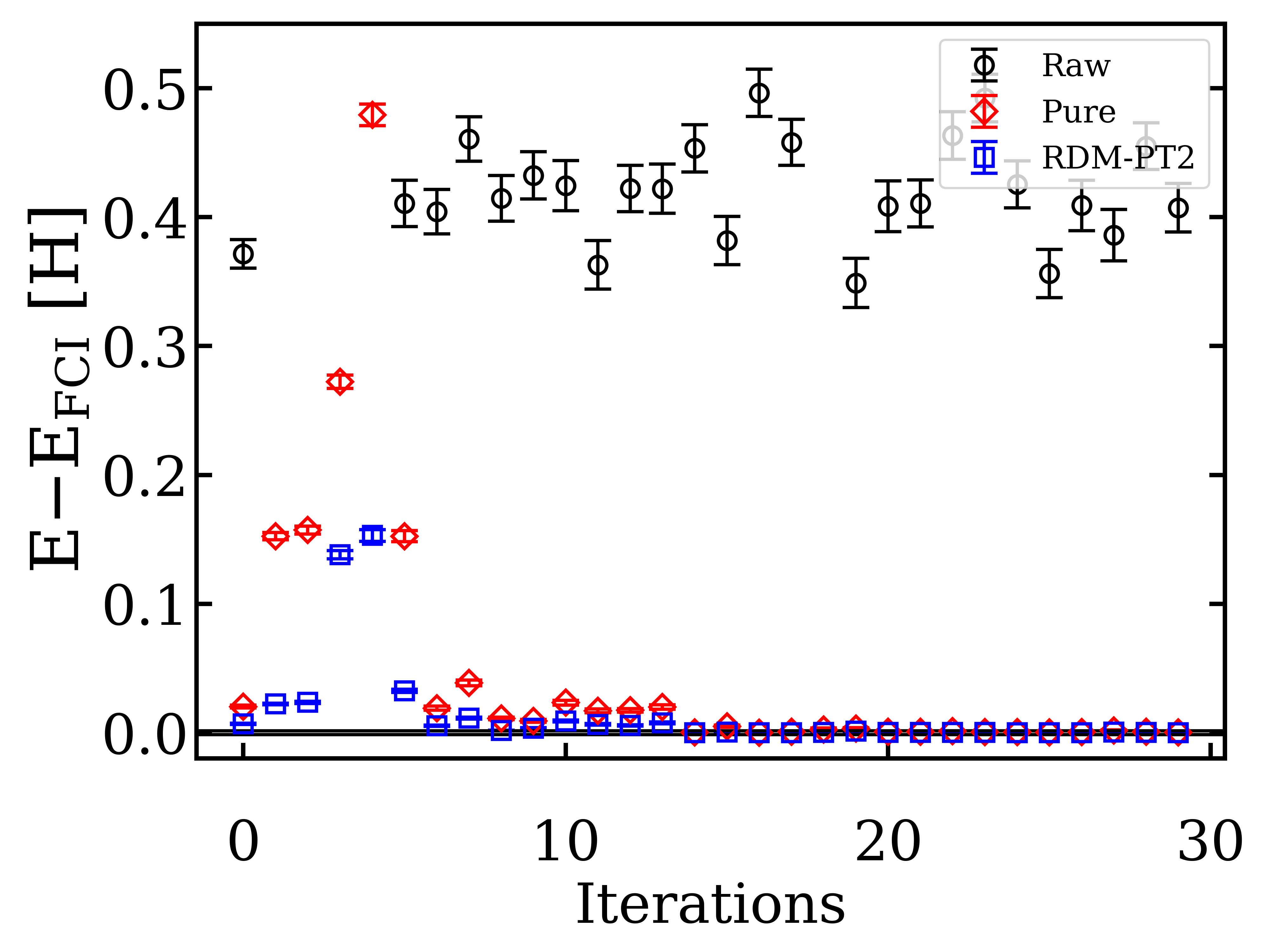}
  \caption{Optimization of ansatz for H$_2$ at the equilibrium geometry of 0.7 angstroms.  }
  \label{fig:h2_opt_eq}
\end{figure}

{\it Results.---}Figure~\ref{fig:h2_opt_eq} shows the optimization of the H$_2$ molecule close to equilibrium geometry of $r=0.7$ angstroms, using the COBYLA optimizer found in scipy.  Plotted are the differences in calculated energy from the exact energy for raw data with only readout error mitigation, pure energies generated from a physical purified density matrix, and perturbatively corrected RDM-PT2 energies. The optimizer searched for optimal parameters based on the purified energies, as the perturbative estimate is not variational and would not be appropriate for the search. 

\begin{figure}[htp]
    \centering
    \includegraphics[width=\columnwidth]{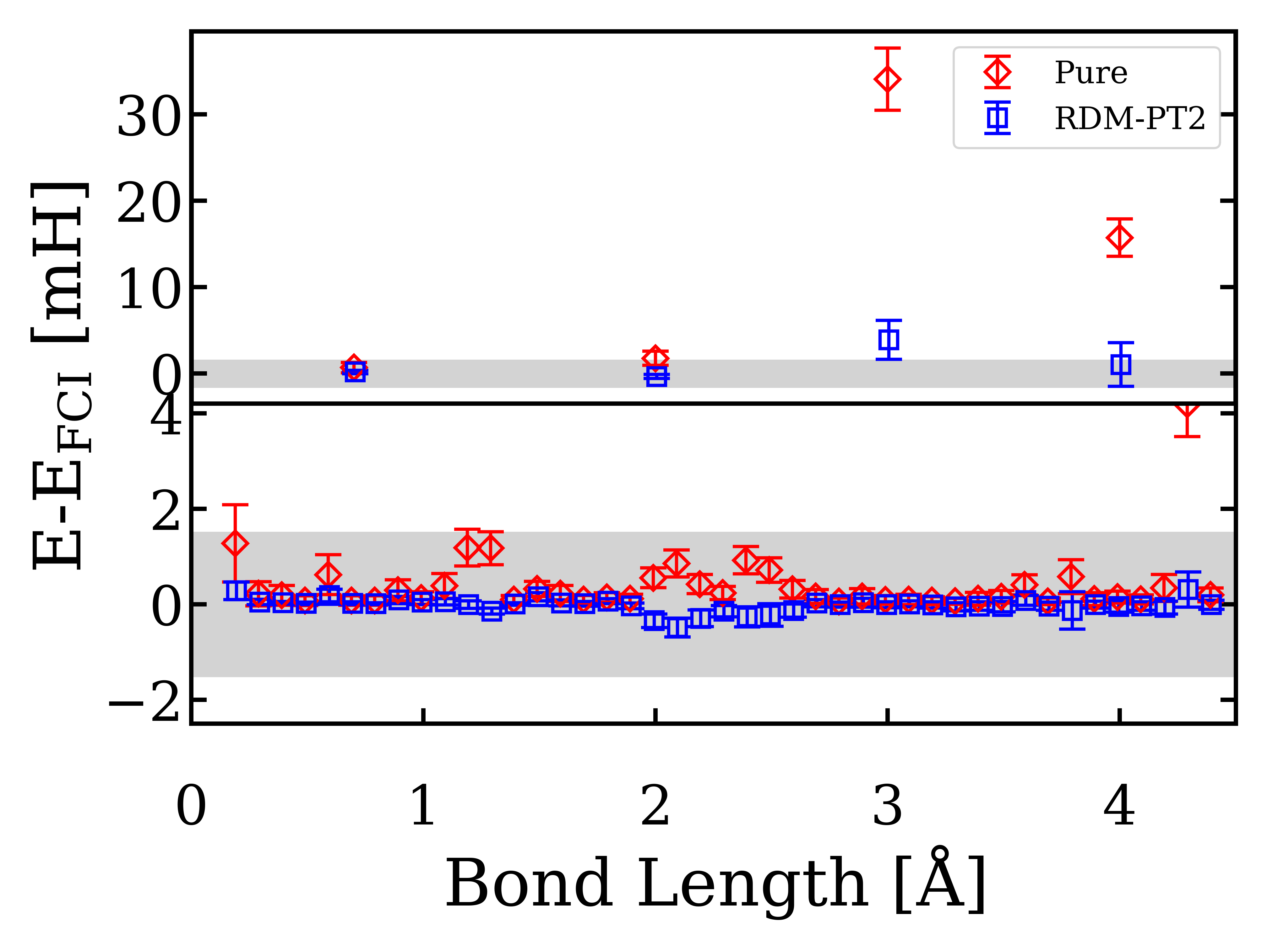}
  \caption{The top panel shows the results of four different quantum simulations run on IBM poughkeepsie, the bottom shows results using the built in Qiskit error simulator with properties obtained from poughkeepsie. }
  \label{fig:h2_dissociation}
\end{figure}

In the top panel of Figure \ref{fig:h2_dissociation}, the pure and RDM-PT2 energy differences measured on IBM quantum hardware are shown for four bond lengths of $r=$0.7,2.0,3.0,and 4.0.  Below are the results from simulated results using the error model taken from the Qiskit module Aer using numbers reported as realistic on IBM Poughkeepsie.  Final results and errors for each geometry were estimated by taking the average and standard deviation of the last 5 points of the COBYLA optimization, and adding it to statistical error in quadrature. For the simulated results from Aer, purified results already fall within
the chemically accurate band of 1.5 milliHartree of the exact result, but RDM-PT2 always yields an improvement. For the calculations on quantum hardware, one sees a different story.  At larger bond lengths, it was more difficult to achieve adequate agreement with exact energies with only purification.  However, the RDM-PT2 energies were able to reach exact results, within error bars.  It is important to note that if the approximately transformed energies are not used as the bond length increases, the perturbative estimate dramatically overestimates the needed correction, yielding overbound results.  It will be interesting to see how this correction fairs in more complicated dissociation curves that involve double or triple bond breaking, where degeneracy often spoils perturbative corrections.

\begin{figure}[htp]
    \centering
    \includegraphics[width=\columnwidth]{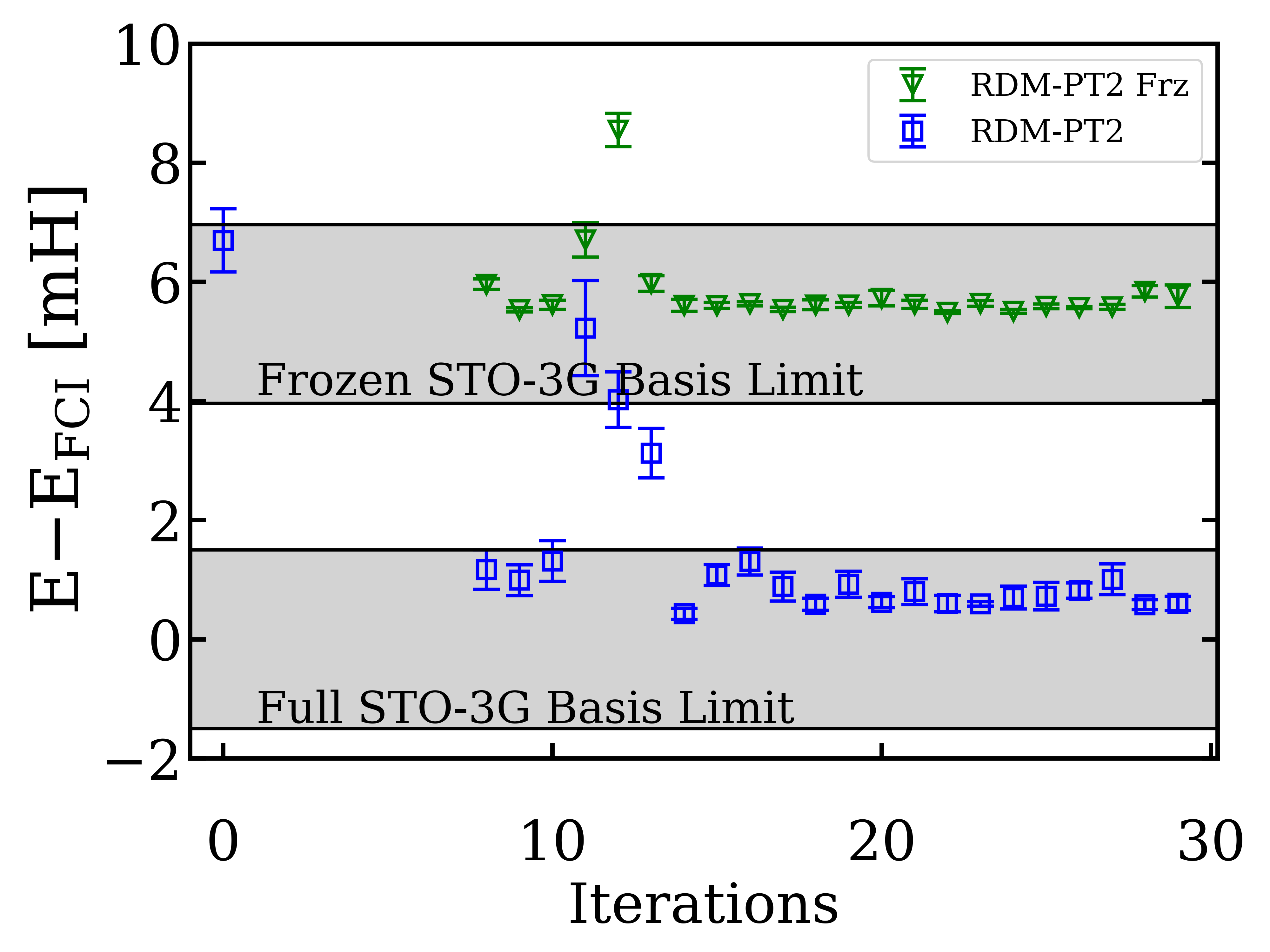}
  \caption{Equlibrium simulation of LiH, shown are the two bands corresponding to exact energies within the frozen and full space sto-3g spaces.}
  \label{fig:LiH_eq_unfrz}
\end{figure}

\begin{figure}[htp]
    \centering
    \includegraphics[width=\columnwidth]{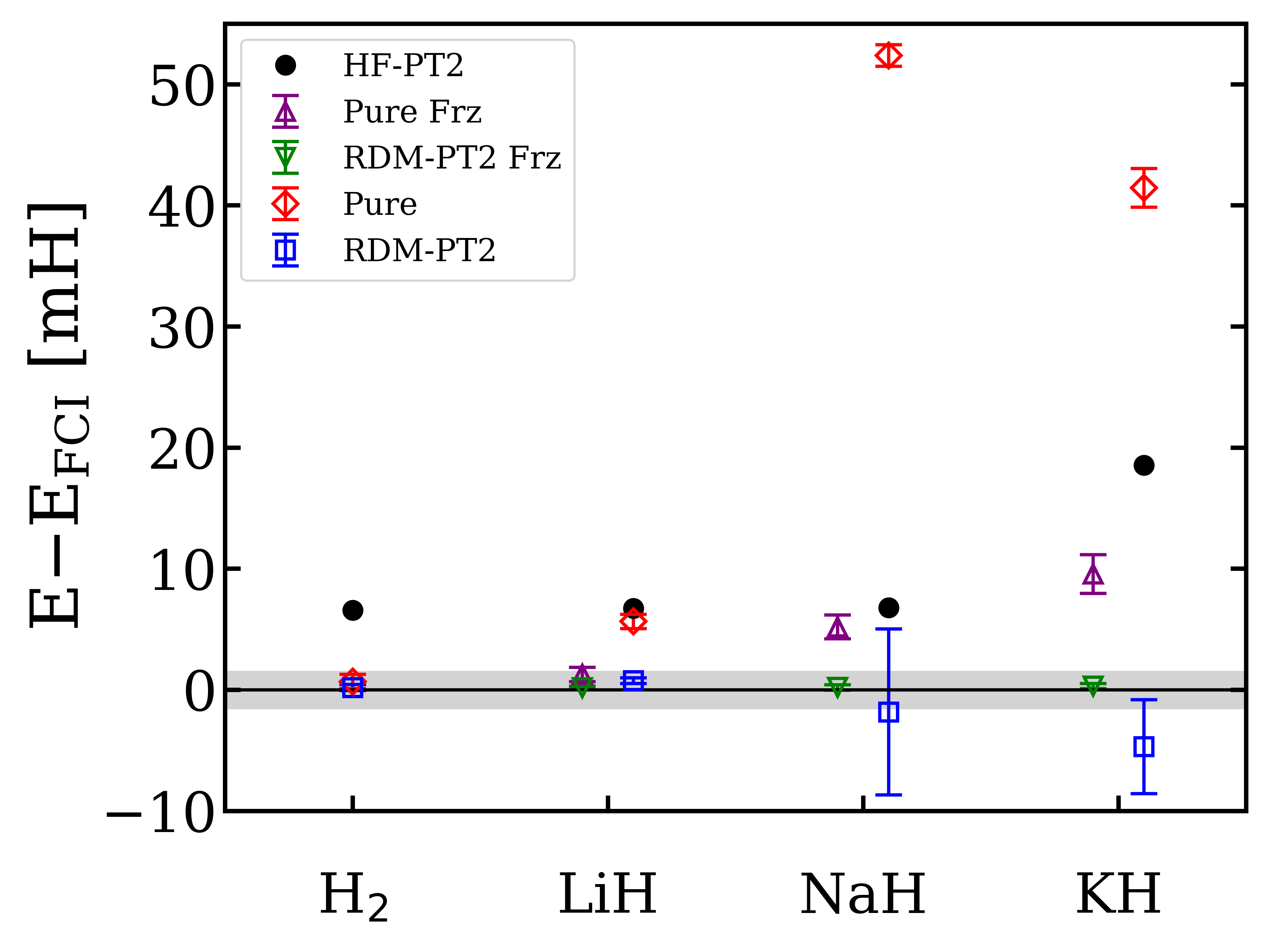}
  \caption{Final results obtained from hardware simulations run on IBM hardware for equilibrium geometries of frozen core hydride type molecules. Although purified results show great variability with regards to machine and individual run, the perturbative expression seems fairly robust.}
  \label{fig:hydrides}
\end{figure}

Figure~\ref{fig:LiH_eq_unfrz} shows the optimization on hardware of LiH at its equilibrium geometry.  The top band shows chemical accuracy with respect the frozen space, while the bottom band is with respect to the full space.  It is clear that the RDM-PT2 using only frozen orbitals and the RDM-PT2 where all orbitals are correlated both give very good agreement with their respective spaces.  Figure~\ref{fig:hydrides} reflects the same analysis for all the hydrides, H$_2$, LiH, NaH, and KH.  For the hydrides heavier than H$_2$, the frozen core energy deviations are shown with a slight left offset, while the full space deviations are slightly offset to the right.  In all cases, it was possible to achieve final RDM-PT2 energies whose error bars fell within 1.5 mHa of the exact energy.  Given for reference are the second order many-body perturbation theory energies with respect to a hartree-fock reference.  Our perturbative corrections are not just reproducing naive perturbation theory, they achieve chemical accuracy whereas HF-PT2 energies do not.  


{\it Conclusion.---}We remotely performed quantum computation of simple hydride molecules using the minimal basis set.  Using the standard process of freezing all but the most important orbitals, these problems were simulable on
quantum hardware available from IBM via the cloud. More importantly, we derived a new, but simple perturbative correction that can augment current error mitigation techniques to dramatically improve ground state measurements of variational quantum eigensolvers.  In addition to bringing the small space calculations we performed into very good agreement with exact energies, the correction also allows for the unfreezing of dynamically-correlated orbitals.  This opens up a path for gradual improvement of calculations as quantum computers grow into larger systems, a key requirement to eventually reach quantum advantage.

{\it Acknowledgments}
This work was supported as part of the ASCR Quantum Testbed Pathfinder Program at Oak Ridge National Laboratory under FWP \#ERKJ332.  This research used quantum computing system resources of the Oak Ridge Leadership Computing Facility, which is a DOE Office of Science User Facility supported under Contract DE-AC05-00OR22725. Oak Ridge National Laboratory manages access to the IBM Q System as part of the IBM Q Network.

\bibliographystyle{apsrev4-1}
\bibliography{references}
\clearpage
\newpage
\appendix 
\setcounter{equation}{0}
{\it Supplemental Information}
In order to motivate the perturbative expansion about the ground state energy, we write down the energy expression for a differential of the energy with respect to a given cluster amplitude, $T_I$ with excitation operator $A_I$, which for reasons that will become apparent, we will call the "middle" energy differential 
\begin{equation}\label{eq:middle_derive}
    \Delta E_{M,I} = \langle\phi|e^{-(T-T^\dagger)-\Delta_I A_I}He^{T-T^\dagger+\Delta_I A_I}|\phi\rangle \,.
\end{equation}
Additionally, we define two more differentials corresponding to the "inner" and "outer" exponentials, i.e.
\begin{equation}\label{eq:inner_derive}
    \Delta E_{I,I} = \langle\phi|e^{-(T-T^\dagger)}e^{-\Delta_I A_I}He^{\Delta_I A_I}e^{T-T^\dagger+\Delta_I A_I}|\phi\rangle \,,
\end{equation}
and
\begin{equation}\label{eq:outer_derive}
    \Delta E_{O,I} = \langle\phi|e^{-\Delta_I A_I}e^{-(T-T^\dagger)}He^{T-T^\dagger}e^{\Delta_I A_I}|\phi\rangle \,.
\end{equation}
We now observe that if one carefully expands all three above expressions, they are all identical up to expressions involving the
three commutator expressions (and higher commutator powers of $T-T^\dagger$, which we neglect to discuss here as we expect they are negligible)
$[T-T^\dagger,[\Delta_I A_I,H]]$,$[\Delta_I A_I[T-T^\dagger,H]]$, and $[H,[\Delta_I A_I,T-T^\dagger]]$.  Since $A_I$ is one given excitation, 
and finite rank truncated Unitary Coupled Cluster is suspected to only be valid if the cluster amplitudes required are not large, it is likely
that all three commutators are vanishingly small as the solution is approached.  This claim would likely not be valid for large amplitudes, however in 
the cases inspected here it appears this is true given the success presented.  Thus we make the approximation $\Delta E_{M,I}\approx\Delta E_{I,I}\approx\Delta E_{O,I}=\Delta E_{I}$, where we will now dispense with the subscript corresponding to "outer", "middle", and "inner".  
It is then possible to take the differential limit,  and make the identification of $\bar{f}_{ia}=\frac{1}{2} \frac{\delta E_{ia}}{\delta T_{ia}}$ and $\bar{\Gamma}_{ijab}=\frac{1}{2}\frac{\delta E_{ijab}}{\delta T_{ijab}}$ with our desired "transformed" matrix elements for both 
single and double excitations out of the Slater Determinant $|\phi\rangle$ and approximate .  In order to evaluate these matrix elements, we use the expressions
that depend only on the 2-RDM that come from Eq. \ref{eq:derive_general}. These are: 
\begin{align}
&\bar{f}_{ia} =  \Re(\langle \phi | [a^\dagger_i a_a,\bar{H}]|\phi \rangle) \approx \Re(\langle\phi|U^\dagger[a^\dagger_i a_a,H]U|\phi\rangle)\notag \\ 
&= \sum_m \big( h_{im}\rho_{ma}-h_{am}\rho_{mi}\big) \notag \\
&+\frac{1}{2}\sum_{m,v,w}\big(g_{imvw}\rho_{vwam}-g_{amvw}\rho_{vwim}\big) \label{eq:onebodygrad}
\end{align}

\begin{align}
&\bar{\Gamma}_{ijab} =  \Re(\langle \phi | [a^\dagger_ia^\dagger_j a_b a_a,\bar{H}]|\phi \rangle) \notag\\
&\approx \Re(\langle\phi|U^\dagger[a^\dagger_ia^\dagger_j a_b a_a,H]U|\phi\rangle)\notag \\
&= (1-P_{ij})\sum_m h_{im}\rho_{mjab}
-(1-P_{ab})\sum_m h_{am}\rho_{ijmb} \notag \\
&+\frac{1}{2}\sum_{m,n}g_{ijmn}\rho_{mnab}
-\frac{1}{2}\sum_{m,n}g_{mnab}\rho_{ijmn}\notag \\
&-(1-P_{ij})\frac{1}{2}\sum_{m,n,v}g_{ivmn}\rho_{mnjabv}\notag\\ 
&+(1-P_{ab})\frac{1}{2}\sum_{m,n,v}g_{mnav}\rho_{ijvbmn}\label{eq:twobodygrad}
\end{align}

The 3-body dependence can be reduced to two-body dependence with the following formula, which is just the "reducible"
3-RDM found in \cite{mazziotti2007reduced}:
\begin{align}
\rho_{pqrstu} \approx \frac{1}{3}(1-P_{pr}-P_{qr})(1-P_{su}-P_{tu})\rho_{pqst}\rho_{ru} \label{eq:3RDM}
\end{align}

In order to better approximate the transformed energy denominators, we appeal to the fact that for naive many-body perturbation theory
power counting, the leading order off-diagonal density matrices can be related to the cluster amplitudes, i.e.
\begin{align}
\rho_{ia}^{[1]} =  T_{ia}^{[1]} \\ 
\rho_{ijab}^{[1]} =  T_{ijab}^{[1]} \,.
\end{align}
We make use of this to form a very rough approximation of the cluster amplitudes in order to approximate the diagonal energies by 
making corrections from the leading order contractions of the cluster amplitudes with the normal ordered hamiltonian.
This gives transformed energies that again depend only on the original Hamiltonian and the measured two body density matrices as follows:
\begin{align}
\epsilon_{i} = h_{ii}+\sum_{\substack{j}}g_{ijij}-\sum_a h_{ia}\rho_{ai}-\frac{1}{2}\sum_{\substack{j,a,b}}g_{ijab}\rho_{abij}\notag\\ \label{eq:holespe} \\
\epsilon_{a} = h_{aa}+\sum_{\substack{j}}g_{ajaj}+\sum_i h_{ai}\rho_{ia}+\frac{1}{2}\sum_{\substack{i,j,b}}g_{abij}\rho_{ijab}\notag\\ \label{eq:partspe}
\end{align}
\end{document}